\documentclass[12pt]{iopart}

\usepackage{iopams}
\usepackage{graphicx}
\usepackage{color}
\usepackage{mathrsfs}
\usepackage{bm}

\begin{document}

\title{Guiding-center kinetic model based on the assumption of homogeneous distribution over gyrophase}

\author{Shuangxi Zhang }


\address{
 Graduate School of Energy Science, Kyoto University, Uji, Kyoto 611-0011, Japan. }


\ead{zhang.shuangxi.3s@kyoto-u.ac.jp}

\date{<date>}

\begin{abstract}
The purpose of this paper is to develop a simplified model as the modeling of the magnetized plasmas. The starting point is an assumption that the distribution of the ensemble of charged particles in the same species is homogeneous over gyrophase. The particles in this ensemble are located at the same guiding-cneter position $(\mathbf{X},\mu,U)$. Then, a fundamental Lagrangian differential 1-form is developed. It contains all particles in the magnetized plasma system as well as the Coulomb pair force between particles instead of field-particle interaction used in conventional gyrokinetic models. By transforming the Lagrangian 1-form to the new one on guiding-center coordinate with the ensemble summation over gyrophase, the new fundamental 1-form is naturally independent of gyrophase of each particle based on the homogeneous distribution over gyrophase, and it determines the dynamics of all particles on the new coordinates. By using a coarse-grained scheme, this new 1-form can be modeled by the guiding-center kinetic model.
\end{abstract}



\maketitle

\section{Introduction}\label{sec1}

The composited coordinate transform used in conventional gyrokinetic models(CGM)\cite{1983wwlee,1983dubin,1988Hahm,1990brizard,2000sugama,2010garbet,2007brizard1,2000qinhong1,2010scott,1983littlejohn,2006shaojiewang} comprises two independent but consecutive coordinate transform. The first one is guiding center transform\cite{1983dubin,1988Hahm,1990brizard}; the other one is gyrocenter transform\cite{1983dubin,1988Hahm,1990brizard,2000sugama}. The purpose of this composited transform is to reduce the gyrophase from the orbit equation of the charged particles, so that the equation of the evolution of the distribution function only solves a five-dimensional distribution\cite{wwleejcp1987,1995lin,2008idomura,2009peeters,2000jenko,2003chenyangjcp,2007jolliet,yuzhipop2009}. The Poisson equation is solved on particle coordinates, so that the distribution needs to be transformed back to the one on particle coordinates to solve the Poisson equation.

In this paper, a simplified model as the modeling of magnetized plasmas is developed. The starting point is the assumption that  the distribution of the ensemble of charged particles in the same species is homogeneous over gyrophase. The particles in this ensemble are located at the same guiding-cneter position $(\mathbf{X},\mu,U)$. This assumption could lead to a significant simplification of the gyrokinetic model through the following steps.

Firstly, the ensemble of particle in the same species and of the same guiding-center coordinate $(\mathbf{X},\mu,U)$ but with their $\theta$ homogeneously distributed over $(0,2\pi]$, can be treated as an ensemble of identical particles, which have the same guiding-center orbit.

Secondly, instead of using the Lagrangian differential 1-form of a test particle as down in CGM \cite{1988Hahm,1990brizard}, this paper implements a
fundamental Lagrangian 1-form which determines the dynamics of all ions and electrons on particles' coordinates.
The electrostatic potential in this Lagrangian is originated from the mutual interactions between charged particle pairs.

Thirdly, according to Lie transform perturbation theory, by carrying out the pullback transform over this Lagrangian 1-form, a new Lagrangian 1-form on guiding-center coordinate is derived, with all gyrophase cancelled by the summation of the gyrophase for each particle included in the ensemble surrounding the guiding-field magnetic field line. This kind of summation will be called ensemble summation in this paper. Therefore, it doesn't need to carry out an additional gyrocenter transform to reduce the gyrophase for each particle, which nevertheless are mutually cancelled by the ensemble summation.

Forthly, the new fundamental Lagrangian 1-form determines the dynamics of all ions and electrons on guiding-center coordinates. Then,  new guiding-center kinetic models(GCKM) can be derived based on a  coarse-grained scheme as the modeling of the new fundamental Lagrangian 1-form.

%

The rest of this paper is arranged as follows. In Sec.(\ref{sec3}), the fundamental Lagrangian 1-form determining the dynamics of all ions and electrons is introduced and is modeled by the Distribution-Poisson models. The modeling procedure will be used to derive new GCKM on guiding center coordinates. In Sec.(\ref{sec4}), the new 1-form on the new coordinates approximated up to the second order is derived by pulling the fundamental Lagrangian 1-form on particle's coordinates back to the one on the new coordinates. In Sec.(\ref{sec5}), with the same modelling method given by Sec.(\ref{sec3}), the new 1-form on the new coordinates with second order approximation is modeled by new GCKM. In Sec.(\ref{sec6}), the equation for the quasi-neutral condition is introduced on guiding-center coordinate. Sec.(\ref{sec7}) is the simple introduction of the numerical application of this new model. Sec.(\ref{sec8}) is dedicated to summary and discussion.


\section{Modelling the fundamental Lagrangian 1-form on particle's coordinates by Distribution-Poisson models}\label{sec3}


It's well-known that the dynamics of a physical system can be determined by the Lagrangian of this system \cite{2006marsden, 1995peskin,1989arnoldbook}. The force experienced by each entity in the system is given by the various potentials in the Lagrangian. In this paper, we focus on the electrostatic plasma including only one species of ion and electrons. In real physical systems, particles are located at different spatial positions. The electrostatic potential experienced by one charged particle with spatial coordinate $\mathbf{x}$ is
\begin{equation}\label{ef131}
\phi \left( {{\bf{x}},t} \right) = \frac{1}{{4\pi {\epsilon_0}}}\sum\limits_j' {\left( {\frac{q}{{|{\bf{x}} - {{\bf{x}}_{ij}}(t)|}} - \frac{e}{{|{\bf{x}} - {{\bf{x}}_{ej}}(t)|}}} \right)},
\end{equation}
where the summation is taken for all ions and electrons except the one whose coordinate is $\mathbf{x}$, as the superscript $'$ indicates.

The fundamental Lagrangian 1-form is the summation of the Lagrangian 1-form for each electron and ion. Here, we only consider the electrostatic case. The magnetic field is the background field independent of all the particles. The case of electromagnetic perturbations, for which the perturbed magnetic field can not be treated as background field, will be considered in future work. The Lagrangian can be written as the summation of two parts as
\begin{equation}\label{ef35}
\gamma  = {\gamma _i} + {\gamma _e},
\end{equation}
 with
\begin{equation}\label{ef14}
{\gamma _i} = \sum\limits_j {\left[ {\begin{array}{*{20}{l}}
{\left( {q{\bf{A}}\left( {{{\bf{x}}_{ij}}} \right) + {m_i}{{\bf{v}}_{ij}}} \right)\cdot d{{\bf{x}}_{ij}}}\\
{ - \left( {\frac{{{m_i}{\bf{v}}_{ij}^2}}{2} + \frac{{q\phi \left( {{{\bf{x}}_{ij}},t} \right)}}{2}} \right)dt}
\end{array}} \right]},
\end{equation}
\begin{equation}\label{ef15}
{\gamma _e} = \sum\limits_j {\left[ {\begin{array}{*{20}{l}}
{\left( { - e{\bf{A}}\left( {{{\bf{x}}_{ej}}} \right) + {m_e}{{\bf{v}}_{ej}}} \right)\cdot d{{\bf{x}}_{ej}}}\\
{ - \left( {\frac{{{m_e}{\bf{v}}_{ej}^2}}{2} - \frac{{e\phi \left( {{{\bf{x}}_{ej}},t} \right)}}{2}} \right)dt}
\end{array}} \right]}.
\end{equation}
The reason for the factor $\frac{1}{2}$ in $\frac{{\phi \left( {{{\bf{x}}_{oj}}} \right)}}{2}$ is that the potential between each pair has mutual contributions by the two particles.
Eq.(\ref{ef15}) can also be written as compactly
\begin{equation}\label{ef36}
\gamma  = \sum\limits_{o \in \{ i,e\} } {\sum\limits_j {\left[ {\begin{array}{*{20}{l}}
{\left( {{q_o}{\bf{A}}\left( {{{\bf{x}}_{oj}}} \right) + {m_o}{{\bf{v}}_{oj}}} \right)\cdot d{{\bf{x}}_{oj}}}\\
{ - \left( {\frac{{{m_o}{\bf{v}}_{oj}^2}}{2} + \frac{{{q_o}\phi \left( {{{\bf{x}}_{oj}},t} \right)}}{2}} \right)dt}
\end{array}} \right]} }.
\end{equation}
$o\in\{i,e\}$ and  $q_i=q$ and $q_e=-e$ are utilized here and will be adopted throughout this paper.

The Lagrangian given by Eq.(\ref{ef36}) determines the dynamics of all ions and electrons included in this system. It denotes  a completely autonomous system.  The equations of motion for each ion and electron can be derived based on Euler-Lagrange equation for each pair of $(\mathbf{x}_{(i,e)j},\mathbf{v}_{(i,e)j})$. The electrostatic potential originates from the mutual interaction of each pair of charged particles. However, it's nearly impossible to straightforwardly calculate the electrostatic potential based on mutual interaction, since there are so many particle pairs. Noting that the electrostatic force generated by each charged particle is of the inverse-square force form, it's well-known that based on Gauss's law, Poisson's equation can directly associate the potential at a spatial point with the charge at that point as follows\cite{1999jackson}
\begin{equation}\label{vp54}
{\nabla ^2}\phi \left( {\bf{x}} ,t\right) =  - \frac{1}{{{\epsilon_0}}}\sum\limits_j {\left[ {q\delta \left( {{\bf{x}} - {{\bf{x}}_{ij}}}(t) \right) - e\delta \left( {{\bf{x}} - {{\bf{x}}_{ej}}}(t) \right)} \right]}.
\end{equation}
Therefore, Poisson's equation plus a boundary condition can be an alternative way to calculate the electrostatic potential particles feel. Eq.(\ref{vp54}) is rigorously based on the inverse-square force.
If the force generated by the charges changes, this formula also changes, e.g., if mutual interaction force is the magnetostatic force, the equation relating magnetic potential to the current is the Ampere's Law. This will be used later to model the fundamental Lagrangian on guiding center by GCKM.

The solution of $\phi(\mathbf{x},t)$ given by Eq.(\ref{vp54}) can be solved by integrating both sides of Eq.(\ref{vp54}) for each charged particle
with a given boundary condition. However, straightforward applications of Eq.(\ref{vp54}) are almost impossible since there are so many particles in a real plasma system. An effective Distribution-Poisson model is used to replace the fundamental Lagrangian given by Eq.(\ref{ef36}).

\begin{figure}[htbp]
\centering
\includegraphics[height=4cm,width=5cm]{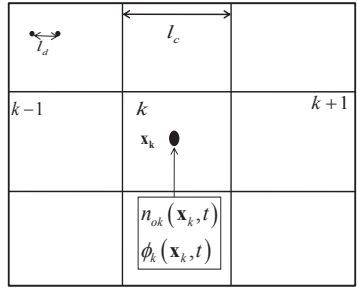}\centering
\caption{\label{cell}  The schematic plot of cells and the location of $n_{ok}(\mathbf{x}_k,t)$ and $\phi_{k}(\mathbf{x}_k,t)$ with $l_d\ll l_c \ll l_p$.}
\end{figure}

It's well-known that to study the phase transition phenomena presented by the Ising model, single spins are replaced by cells containing multiple spins
\cite{1966kadanoff,1982baxter,1971wilson}, since the correlation length of the spin fluctuation is much longer than the distance between two neighbouring spins at the phase transition point. As for the electrostatic potential generated by the accumulation of charges as described by Eq.(\ref{vp54}), if the length scale of electrostatic potential denoted as $l_p$ is much longer than the mean distance between charged particles denoted as $l_d$, to determine the electrostatic potential,
it's not needed to know the specific position of each particle. With the same treatment of the Ising model, a coarse graining process is possible to divide the spatial space occupied by the plasma into small cells with $l_p\gg l_c \gg l_d$ satisfied, where $l_c$ is the size of the cell, as shown in Fig.(\ref{cell}). To determine the electrostatic potential through Poisson's equation given in Eq.(\ref{vp54}), we only need to know the accumulation of charges within each cell. For the theoretical analysis, $l_c$ can be much smaller than the length scale of the grid used in the Particle-In-Cell simulation\cite{birdsallbook}.

Then, the spatial integration of Eq.(\ref{vp54}) is given by the following formula
\begin{equation}\label{vp55}
\begin{array}{l}
\int_V {{\nabla ^2}\phi \left( {{\bf{x}},t} \right){d^3}{\bf{x}}}   \\
= - \frac{1}{{{\epsilon_0}}}\int_V {\sum\limits_{j'} {\left[ \begin{array}{l}
q\delta \left( {{\bf{x}} - {{\bf{x}}_{ij}}(t)} \right)\\
 - e\delta \left( {{\bf{x}} - {{\bf{x}}_{ej}}(t)} \right)
\end{array} \right]} {d^3}{\bf{x}}},
\end{array}
\end{equation}
which can be modeled by the following summation of the quantities within each cell
\begin{eqnarray}\label{vp56}
&&\sum\limits_k {{\nabla ^2}{\phi _k}\left( {{{\bf{x}}_k},t} \right)\Delta {V_k}}  \nonumber \\
&& =  - \frac{1}{{{\epsilon_0}}}\sum\limits_k {\frac{{{\beta _{ik}}(t) - {\beta _{ek}}(t)}}{{\Delta {V_k}}}\Delta {V_k}}.
\end{eqnarray}
$\mathbf{x}_k$ is the coordinate of the spatial center of the $k$th cell. $\phi_k(\mathbf{x}_k,t)$ is the average electrostatic potential of the $k$th cell and is located at the spatial center of the cell. So the gradient operator $\nabla$ over $\phi_k(\mathbf{x}_k,t)$ can be modeled by the middle-point difference. $\Delta V_k$ represents the volume of the $k$th cell. $\beta_{ik}$ and $\beta_{ek}$ are the charge accumulation for ions and electrons within the $k$th cell, respectively. Upon the length scale $l_c$, we make the following replacement
\begin{equation}\label{vp57}
 \frac{{\beta_{ok} }(t)}{{\Delta {V_k}}}\Delta {V_k} = q_o n_{ok}(\mathbf{x}_k,t)\Delta {V_k}.
\end{equation}
$n_{ok}(\mathbf{x}_{k},t)$ is the average density of the $k$th cell for ions or electrons and located at the center of the $k$th cell. The location of the average electrostatic potential and average density is given by Fig.(\ref{cell}).
The location of $\phi_k(\mathbf{x}_k,t)$ and $n_{ok}(\mathbf{x}_k,t)$ is different from that in Particle-In-Cell simulation, where the physical quantities such as the potential and charge density are all located at grid points\cite{birdsallbook}. We take the following approximation of Poisson's equation for the $k$th cell
\begin{equation}\label{vp58}
{\nabla ^2}{\phi _k}({{\bf{x}}_k},t) = \frac{1}{{{\epsilon_0}}}\left( e{{n_{ek}}({{\bf{x}}_k},t) - q{n_{ik}}({{\bf{x}}_k},t)} \right).
\end{equation}
This is the wanted edition of Poisson's equation as the result of coarse graining. The cell's length scale $l_c$ can shrink to be close to $l_d$, so that the approximate potential can be infinitely close to the real potential. For the $j$th particle located within the $k$th cell, the relation between the real potential $\phi \left( {{{\bf{x}}_j}},t \right)$ and the approximate one ${\phi _k}(\mathbf{x}_k,t)$  satisfies the following formula
\begin{equation}\label{ef25}
\phi \left( {{{\bf{x}}_j}},t \right) = {\phi _k}(\mathbf{x}_k,t) + O({l_c}).
\end{equation}

The time evolution of $n_k(\mathbf{x}_k, t)$ is needed. $n_k(\mathbf{x}_k, t)$ can be expressed as an integral of distribution function. The distribution of particles can be given by the Klimontovich  distribution\cite{1982klimontovich}
\begin{equation}\label{ef2}
{M_{ok}}\left( {\bf{z}} \right) = \sum\limits_j {\delta \left( {{\bf{x}} - {{\bf{x}}_{oj}}(t)} \right)\delta \left( {{\bf{v}} - {{\bf{v}}_{oj}}\left( t \right)} \right)}
\end{equation}
The average density $n_{ok}$ within the $k$th cell is given by an integral
\begin{equation}\label{ef3}
{n_{ok}}({{\bf{x}}_k},t) = \frac{1}{{\Delta {V_k}}}\int_{\Delta {V_k}} {{M_{ok}}\left( {\bf{z}} \right){d^3}{\bf{x}}d^3\mathbf{v}}.
\end{equation}


The evolution of ${M}_{ok}\left( \mathbf{z} \right)$ is based on the equations of motion of charged particle, which is in turn determined by the Lagrangian 1-form. The electrostatic potential all the particles feel in the $k$th cell is approximated to be the same  as ${\phi _k}(\mathbf{x}_k,t)$. According to Eq.(\ref{ef25}), by shrinking $l_c$ to be much smaller than the Larmor radius of ion or electron, we could use ${\phi _k}(\mathbf{x}_k,t)$ to replace the real potential $\phi(\mathbf{x}_j,t)$ felt by the $j$th particle at the $k$th cell. As discussed in Appendix.(\ref{sec10}), the magnetic field is the background field  independent of the motion of charged particles and is given in advance. So each particle experiences different magnetic field, since each particle in the $k$th cell has its own position. Then, the test Lagrangian 1-form describing the motion of the $j$th single particle for ions or electrons within the $k$th cell can be extracted out from the fundamental Lagrangian 1-form in Eq.(\ref{ef36}) as
\begin{eqnarray}\label{ef5}
{\gamma _{oj}}&& = \left( {{q_o}{\bf{A}}\left( {{{\bf{x}}_{oj}}} \right) + {m_o}{{\bf{v}}_{oj}}} \right)\cdot d{{\bf{x}}_{oj}}  \nonumber \\
&& - \left( {\frac{{{m_o}{\bf{v}}_{oj}^2}}{2} + {q_o}{\phi _k}({{\bf{x}}_k},t)} \right)dt.
\end{eqnarray}
Here, it should be noted that the factor $\frac{1}{2}$ before the electrostatic potential is removed.
Eq.(\ref{ef5}) gives the trajectory equations of particles at the $k$th cell with the approximate potential $\phi_k(\mathbf{x}_k,t)$.
Collisions, the source and the sink are neglected in this paper. So the evolution of ${M}_{ik}\left( \mathbf{z} \right)$ is governed by the Liouville  equation
\begin{equation}\label{ef6}
\left( {\frac{\partial }{{\partial t}} + \frac{{d{\bf{x}}}}{{dt}}\cdot\nabla  + \frac{{d{\bf{v}}}}{{dt}}\cdot\frac{\partial }{{\partial {\bf{v}}}}} \right)M_{ok}\left(\mathbf{z} \right) = 0.
\end{equation}
The obtained $M_{ok}\left(\mathbf{z} \right)$ can be substituted back in Eq.(\ref{ef3}) to get the average charge density of the $k$th cell. According to Klimontovich's theory\cite{1982klimontovich}, the ensemble summation of $M_{ok}\left( \mathbf{z}\right)$ leads to a distribution function $f_{ok}(\mathbf{z})$ as a continuous function  of argument $\bf{z}$. Then, Eq.(\ref{ef6}) changes to be the Vlasov equation
\begin{equation}\label{ef7}
\left( {\frac{\partial }{{\partial t}} + \frac{{d{\bf{x}}}}{{dt}}\cdot\nabla  + \frac{{d{\bf{v}}}}{{dt}}\cdot\frac{\partial }{{\partial {\bf{v}}}}} \right){f_{ok}}\left(
\mathbf{z} \right) = 0.
\end{equation}

Usually, to solve electron's Vlasov equation, the adiabatic approximation of the fluctuation density of electrons can be utilized\cite{1980antonsen} as
\begin{equation}\label{ef33}
{n_e}({\bf{x}},t) = {n_{e0}}({\bf{x}}) + \frac{{e\phi \left( {{\bf{x}},t} \right)}}{{{T_e}}}{n_{e0}}({\bf{x}}),
\end{equation}
where $n_{0e}(\mathbf{x})$ is the equilibrium density for electrons on particle's coordinates. Dividing $f_o(\bf{z})$ into equilibrium and perturbation parts as $f_i(\mathbf{z})=f_{i0}(\mathbf{z})+\tilde{f}_i(\mathbf{z})$, the charge density of ions generated by the equilibrium part is canceled by the equilibrium part of electrons in the Poisson's equation, the rest of which becomes
\begin{equation}\label{ef34}
{\nabla ^2}{\phi _k}(\mathbf{x}_k,t) = -\frac{1}{{{\epsilon_0}}}\left( {\begin{array}{*{20}{l}}
{q\int {_{\Delta {V_k}}{{\tilde f}_{ik}}\left( {\bf{z}} \right)B({\bf{x}})d{\mu _{1i}}d{U_{1i}}} }\\
{ - \frac{{{e^2}\phi_k \left( {{\bf{x}}_k,t} \right)}}{{{T_e}}}{n_{e0}}({\bf{x}},t)}
\end{array}} \right)
\end{equation}

So far, a group of equations comprising Eqs.(\ref{vp58},\ref{ef5},\ref{ef6})forming the Klimontovich-Poisson model or Eqs.(\ref{vp58},\ref{ef5},\ref{ef7}) forming the Vlasov-Poisson model are derived to constitute a close system to model the real fundamental Lagrangian 1-form given by Eq.(\ref{ef36}). The main difference between the two systems is that the former one utilizes a coarse graining scheme to model the electrostatic potential and charge density, and the electrostatic potential is obtained by solving the Poisson's equation.  When the length scale $l_c$ of the cell is small enough, the subscript $k$ of those equations can be deleted, so that these three equations can be treated as defined on continuous spatial space.



As pointed out previously, the modeling is based on the force between charged particles being inverse-square force, so that  Poisson's equation Eq.(\ref{vp58}) can be derived to relate the potential to the charge density. If the force changes, Poisson's equation should be changed accordingly. For example, if the force is of magnetic origin, the equation should be replaced by the Ampere's law correspondingly. With the same principle, we could develop GCKM as the modeling of the fundamental Lagrangian 1-form on guiding-center coordinates.

\section{The fundamental Lagrangian 1-form on guiding-center coordinates}\label{sec4}

The fundamental Lagrangian differential 1-form for all ions and electrons on particle's coordinates is given by Eq.(\ref{ef36}). In this subsection, a pullback transform is adopted to pulling the 1-form back to a new one on guiding-center coordinates with $\theta$ angle reduced from the whole dynamical system up to order $O(\varepsilon_i^2)$ for ion and $O(\varepsilon_e^2)$ for electrons. Here, the ordering parameters are given as ${\varepsilon _i} \equiv \frac{1}{{q{L_0}}}\sqrt {\frac{{2{\mu _{it }}{B_0}}}{{{m_i}}}}$ and ${\varepsilon _e} \equiv \frac{1}{{e{L_0}}}\sqrt {\frac{{2{\mu _{et}}{B_0}}}{{{m_e}}}} $, with $\mu_{it}$ and $\mu_{et}$ the magnetic moment of the thermal velocity for ions and electrons, respectively.
Before carrying out the calculation, the three assumptions are firstly listed as follows.

(1). The first assumption is that the distribution is homogeneous over gyrophase on particle coordinates.

(2). The second one is that  particles' coordinates only experience guiding center transform, which is resulted directly from the fact (1). As will be shown later, only guiding-center transform could help reduce the gyrophase of all particles from the dynamics of the whole system.



\begin{figure}[htbp]
\centering
\includegraphics[height=5cm,width=6cm]{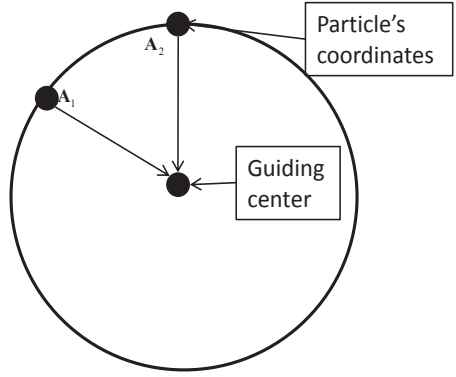}\centering
\caption{\label{remove}  The schematic plot of particles $\mathbf{A}_1$ and $\mathbf{A}_2$, the spatial part of whose new coordinates is located at the same position on new coordinates. The interactions between these particles are removed from the electrostatic potential.}
\end{figure}

(3). The third assumption is associated with removing the singularity faced by the potential function in the new coordinates. After coordinate transform, it's inevitable that the spatial part of some particles' new coordinates could locate at the same spatial position in new coordinate system as shown in Fig.(\ref{remove}). This would introduce singularity into the electrostatic potential, as two identical spatial coordinate appear at the denominate to make it equal zero. To remove the singularity, it's assumed that the mutual potential between those particles located at the same spatial place are removed from the total potential. This assumption is equivalent to the one that the interactions between particles, whose new spatial coordinates after the coordinate transform are the same, are removed from the total potential in Eq.(\ref{ef131}). The removed part occupies only a very small part of the total electrostatic potential.

\subsection{Deriving the fundamental Lagrangian 1-form on the new coordinates by Cary-Littlejohn single-parameter Lie transform method}\label{sec4.1}

With the second assumptions, we can go on to derive the new fundamental Lagrangian 1-form on the new coordinates. As shown in Ref.(\cite{1983cary}), the single-parameter coordinate transform of the coordinates of a single particle is
\begin{equation}\label{gf10}
\frac{{dZ_{ojf}^m}}{{d{\varepsilon _o}}}\left( {{{\bf{z}}_{oj}},{\varepsilon _o}} \right) = g_{oj}^m\left( {{{\bf{Z}}_{oj}}} \right),
\end{equation}
\begin{equation}\label{gf13}
\frac{{d{\bf{z}}_{oj}}}{{d\varepsilon_o }} = 0,
\end{equation}
where not as given previously, $\mathbf{g}_{oj}\left( {{{\bf{Z}}_{oj}}} \right)$ is normalized here with $\mathbf{g}_{oj}\left( {{{\bf{Z}}_{oj}}} \right)\equiv -\bm{\rho}_o/\varepsilon_o$.
The subscript $f$ in Eq.(\ref{gf10}) denotes forward transform. Eq.(\ref{gf13}) can be further written as
\begin{equation}\label{gf14}
\frac{{\partial z_{oj}^i}}{{\partial \varepsilon_o }} + \frac{{dZ_{oj}^k}}{{d\varepsilon_o }}\frac{{dz_{oj}^i}}{{dZ_{oj}^k}} = 0,
\end{equation}
where repeated indexes denote the Einstein summation.
Eqs.(\ref{gf10}) and (\ref{gf14}) induce a coordinate transform
\begin{equation}\label{gf11}
{{\bf{z}}_{oj}}\left( {{{\bf{Z}}_{oj}},{\varepsilon _o}} \right) = \exp \left( { - {\varepsilon _o}g_{oj}^m\left( {{{\bf{Z}}_{oj}}} \right){\partial _{Z_{oj}^m}}} \right){{\bf{Z}}_{oj}}.
\end{equation}

Now, we make following definitions
\begin{equation}
\begin{array}{l}
{\bar {\bf{z}} _o} \equiv \left( {{{\bf{z}}_{o1}},{{\bf{z}}_{o2}},{{\bf{z}}_{o3}}, \cdots } \right),\\
{\bar {\bf{Z}} _o} \equiv \left( {{{\bf{Z}}_{o1}},{{\bf{Z}}_{o2}},{{\bf{Z}}_{o3}}, \cdots } \right),\\
{\bar {\bf{g}} _o} \equiv \left( {{{\bf{g}}_{o1}},{{\bf{g}}_{o2}},{{\bf{g}}_{o3}}, \cdots } \right).  \nonumber \\
\end{array}
\end{equation}
The following backward coordinate transformation can be derived
\begin{equation}\label{gf12}
{\bar {\bf{z}} _{oj}}\left( {{{\bar {\bf{Z}} }_o},{\varepsilon _o}} \right) = \exp \left( { - {\varepsilon _o}\overline g _o^m{\partial _{\bar Z_o^m}}} \right){\bar {\bf{Z}} _o}.
\end{equation}
Now, implementing the method in Ref.(\cite{1983cary}) with Eq.(\ref{gf14}), we first carry out the pullback transform for ions' coordinates. The following equation can be derived
\begin{eqnarray}\label{gf15}
&&\frac{{\partial {\bar{\Gamma} _m}}}{{\partial {\varepsilon _i}}}\left( {{{\overline {\bf{Z}} }_i},{{\overline {\bf{z}} }_e},{\varepsilon _i},{\varepsilon _e}} \right)   \nonumber \\
&&=  - \bar g_i^n\left( {{{\overline {\bf{Z}} }_i}} \right)\left[ {\frac{{\partial {\bar{\Gamma} _m}}}{{\partial {{\bar Z}^n}}}\left( {{{\overline {\bf{Z}} }_i},{{\overline {\bf{z}} }_e},{\varepsilon _i},{\varepsilon _e}} \right) - \frac{{\partial {\bar{\Gamma} _n}}}{{\partial {{\bar Z}^m}}}\left( {{{\overline {\bf{Z}} }_i},{{\overline {\bf{z}} }_e},{\varepsilon _i},{\varepsilon _e}} \right)} \right]  \nonumber \\
&&- \frac{\partial }{{\partial {{\bar Z}^m}}}\left[ {\bar g_i^n\left( {{{\overline {\bf{Z}} }_i}} \right){\bar{\Gamma} _n}\left( {{{\overline {\bf{Z}} }_i},{{\overline {\bf{z}} }_e},{\varepsilon _i},{\varepsilon _e}} \right)} \right],
\end{eqnarray}
which leads to the solution
\begin{equation}\label{gf16}
\bar \Gamma \left( {{{\overline {\bf{Z}} }_i},{{\overline {\bf{z}} }_e},{\varepsilon _i},{\varepsilon _e}} \right) = \exp \left( { - {\varepsilon _o}{L_{{{\overline {\bf{g}} }_i}}}} \right)\gamma \left( {{{\overline {\bf{Z}} }_i},{{\overline {\bf{z}} }_e},{\varepsilon _i},{\varepsilon _e}} \right) + dS.
\end{equation}
Next, the pullback transform for electrons' coordinates is carried out for the 1-form in Eq.(\ref{gf16}), and the following formula is derived
\begin{eqnarray}\label{gf17}
&&\frac{{\partial {{\Gamma} _m}}}{{\partial {\varepsilon _e}}}\left( {{{\overline {\bf{Z}} }_i},{{\overline {\bf{Z}} }_e},{\varepsilon _i},{\varepsilon _e}} \right)  \nonumber \\
&& =  - \bar g_e^n\left( {{{\overline {\bf{Z}} }_e}} \right)\left[ {\frac{{\partial {{\Gamma} _m}}}{{\partial \bar Z_e^n}}\left( {{{\overline {\bf{Z}} }_i},{{\overline {\bf{Z}} }_e},{\varepsilon _i},{\varepsilon _e}} \right) - \frac{{\partial {{\Gamma} _n}}}{{\partial \bar Z_e^m}}\left( {{{\overline {\bf{Z}} }_i},{{\overline {\bf{Z}} }_e},{\varepsilon _i},{\varepsilon _e}} \right)} \right] \nonumber \\
\;\;\;&& - \frac{\partial }{{\partial \bar Z_e^m}}\left[ {\bar{g}_{e}^n\left( {{{\overline {\bf{Z}} }_i}} \right){{\Gamma} _n}\left( {{{\overline {\bf{Z}} }_i},{{\overline {\bf{Z}} }_e},{\varepsilon _i},{\varepsilon _e}} \right)} \right],
\end{eqnarray}
which leads to the solution
\begin{equation}\label{gf18}
\Gamma \left( {{{\overline {\bf{Z}} }_i},{{\overline {\bf{Z}} }_e},{\varepsilon _i},{\varepsilon _e}} \right) = \exp \left( { - {\varepsilon _e}{L_{{{\overline {\bf{g}} }_e}}}} \right)\bar{\Gamma} \left( {{{\overline {\bf{Z}} }_i},{{\overline {\bf{Z}} }_e},{\varepsilon _i},{\varepsilon _e}} \right) + dS.
\end{equation}
By substituting Eq.(\ref{gf16}) in Eq.(\ref{gf18}), the latter becomes
\begin{eqnarray}\label{gf19}
\Gamma \left( {{{\overline {\bf{Z}} }_i},{{\overline {\bf{Z}} }_e},{\varepsilon _i},{\varepsilon _e}} \right) = && \exp \left( { - {\varepsilon _i}{L_{{{\overline {\bf{g}} }_i}}} - {\varepsilon _e}{L_{{{\overline {\bf{g}} }_e}}}} \right)\gamma \left( {{{\overline {\bf{Z}} }_i},{{\overline {\bf{Z}} }_e},{\varepsilon _i},{\varepsilon _e}} \right)  \nonumber \\
&& + dS
\end{eqnarray}
which can be further written as
\begin{equation}\label{ef21}
\Gamma  = \exp \left( { - \sum\limits_j {\left( {{\varepsilon _i}{L_{{\bf{g}}_{ij}^{\bf{X}}}} + {\varepsilon _e}{L_{{\bf{g}}_{ej}^{\bf{X}}}}} \right)} } \right)\gamma \left( {\bf{Z}} \right),
\end{equation}
for the simplicity.  Since each coordinate pair $(\mathbf{x}_{(i,e)j},\mathbf{v}_{(i,e)j})$ is independent of all others, the operators ${L_{{\bf{g}}_{(i,e)j}^{\bf{X}}}}$ for each $j$ commutes.

\subsection{Approximating Eq.(\ref{ef21}) to the order of $O(\varepsilon_o^2)$}\label{sec4.2}

Expanding Eq.(\ref{ef21}) based on the order of $\varepsilon_o$ and $\varepsilon_e$, the eventual Lagrangian differential 1-form, can be derived. The following rules will be used
\begin{equation}\label{ff2}
\begin{array}{l}
{L_{{\bf{g}}_1^{\bf{x}}}}\left( {{\bf{f}}({\bf{Z}})\cdot d{\bf{X}}} \right) =  - {\bf{g}}_1^{\bf{x}} \times \nabla  \times {\bf{f}}\left( {\bf{Z}} \right)\cdot d{\bf{X}}\\
 - {\bf{g}}_1^{\bf{x}}\cdot\left( {{\partial _t}{\bf{f}}({\bf{Z}})dt + {\partial _\theta }{\bf{f}}({\bf{Z}})d\theta  + {\partial _\mu }{\bf{f}}({\bf{Z}})d\mu } \right) + dS,
\end{array}
\end{equation}
\begin{equation}\label{ff3}
{L_{{\bf{g}}_1^{\bf{x}}}}\left( {h({\bf{Z}})dt} \right) = {\bf{g}}_1^{\bf{x}}\cdot\nabla h\left( {\bf{Z}} \right)dt+dS.
\end{equation}
Here, $\mathbf{f}(\mathbf{Z})$ and $h(\mathbf{Z})$ are any vector function and scalar function on the new coordinates, respectively.

Among the expansions, the following 1-form is denoted as $\Gamma_0$
\begin{equation}\label{vp35}
{\Gamma _0} = \sum\limits_{o \in \{ i,e\} } {\sum\limits_j {\left[ {\begin{array}{*{20}{l}}
{\left( {\begin{array}{*{20}{l}}
{\frac{q_o}{\varepsilon_o}{\bf{A}}\left( {{{\bf{X}}_{oj}}} \right) + {m_o}{U_{oj}}{\bf{b}}}\\
{ + \sqrt {\frac{{2B({{\bf{X}}_{oj}}){\mu _{oj}}}}{{{m_o}}}} {{\widehat {\bf{v}}}_ {oj\bot} }}
\end{array}} \right)\cdot d{{\bf{X}}_{oj}}}\\
{ - \left( {\begin{array}{*{20}{l}}
{\frac{{{m_o}U_{oj}^2}}{2} + {\mu _{oj}}B({{\bf{X}}_{oj}})}\\
{ + \frac{{{q_o}\Phi \left( {{{\bf{X}}_{oj}},t} \right)}}{2\varepsilon_o}}
\end{array}} \right)dt}
\end{array}} \right]} }.
\end{equation}
Here, the factor $\varepsilon_o$ for $o \in \{i,e\}$ is used as a symbol to denote the order of the term adjacent to it. This terminology will be used throughout the remaining part of this paper.  The electrostatic potential is
\begin{equation}\label{ef24}
\Phi \left( {{{\bf{X}}_{oj}},t} \right) = \frac{1}{{4\pi {\epsilon_0}}}\sum\limits_h ' \left( \begin{array}{l}
\frac{q}{{|{{\bf{X}}_{oj}} - {{\bf{X}}_{ih}}(t)|}}\\
 - \frac{e}{{|{{\bf{X}}_{oj}} - {{\bf{X}}_{eh}}(t)|}}
\end{array} \right),
\end{equation}
where $'$ denotes all particles located at $\mathbf{X}_{oj}$ are removed from the summation based on the third assumption.
%

Due to the homogeneous property of the distribution of particles over $\theta$, the summation in Eq.(\ref{vp35}) cancels terms depending on the gyroangle, leading to the following formula
\begin{equation}\label{ff29}
{\left. {\sum\limits_l {\sqrt {\frac{{2B\left( {{{\bf{X}}_{ol}}} \right){\mu _{ol}}}}{{{m_o}}}} } {{\widehat {\bf{v}}}_{ol}}\cdot d{{\bf{X}}_{ol}}} \right|_{\scriptstyle{{\bf{X}}_{ol}} = {{\bf{X}}}\hfill\atop
{\scriptstyle{U_{ol}} = {U}\hfill\atop
\scriptstyle{\mu _{ol}} = {\mu }\hfill}}} = 0,
\end{equation}
with ${\widehat {\bf{v}}_{ol}} = {\bf{e}}_1\sin {\theta _{ol}} + {{\bf{e}}_2}\cos {\theta _{ol}}$. Eq.(\ref{ff29}) can be understood in another way
\begin{equation}\label{ff32}
{\left. {\sum\limits_l {\sqrt {\frac{{2B\left( {{{\bf{X}}_{ol}}} \right){\mu _{ol}}}}{{{m_o}}}} } {{\widehat {\bf{v}}}_{ol}}\cdot{{\mathop {\bf{X}}\limits^. }_{ol}}} \right|_{\scriptstyle{{\bf{X}}_{ol}} = {{\bf{X}}}\hfill\atop
{\scriptstyle{U_{ol}} = {U}\hfill\atop
\scriptstyle{\mu _{ol}} = {\mu}\hfill}}} = 0.
\end{equation}
The reason for the standing of Eq.(\ref{ff29}) or Eq.(\ref{ff32}) is as follows. Given any group $\{\mathbf{X},\mu,U\}$, we have a ensemble of ions or electrons, in which the spatial part of guiding-center coordinates, the parallel velocity  and  magnetic moment of each particle equal $\{\mathbf{X},\mu,U$\}, respectively. The particles in this ensemble are homogeneously distributed over $\theta$  surrounding the guiding center $\mathbf{X}$. So, the summation of $\sin\theta_{ol}$ or $\cos\theta_{ol}$ for all $l$ equals zero, where  subscript $l$ denotes particles of the same $\{\mathbf{X},U,\mu\}$.
Then, $\Gamma_0$ becomes
\begin{equation}\label{ff5}
{\Gamma _0} = \sum\limits_{o \in \{ i,e\} } {\sum\limits_j {\left[ {\begin{array}{*{20}{l}}
{\left( {\frac{q_o}{\varepsilon_o}{\bf{A}}\left( {{{\bf{X}}_{oj}}} \right) + {m_o}{U_{oj}}{\bf{b}}} \right)\cdot d{{\bf{X}}_{oj}}}\\
{ - \left( {\begin{array}{*{20}{l}}
{\frac{{{m_o}U_{oj}^2}}{2} + {\mu _{oj}}B({{\bf{X}}_{oj}})}\\
{ + \frac{{{q_o}\Phi \left( {{{\bf{X}}_{oj}},t} \right)}}{2\varepsilon_o}}
\end{array}} \right)dt}
\end{array}} \right]} }.
\end{equation}

The next one is
\begin{eqnarray}\label{ff1}
{\Gamma _1} && =  - \sum\limits_{o \in \{ i,e\} } {\sum\limits_j {{\varepsilon_o L_{{\bf{g}}_{oj}^{\bf{X}}}}\gamma \left( {\bf{Z}} \right)} }  \nonumber \\
&& = \sum\limits_{o \in \{ i,e\} } {\sum\limits_j {\left[ {\begin{array}{*{20}{l}}
{ - \varepsilon_o {L_{{\bf{g}}_{oj}^{\bf{X}}}}\left( {\left( {\begin{array}{*{20}{l}}
{\frac{q_o}{\varepsilon_o}{\bf{A}}\left( {{{\bf{X}}_{oj}}} \right) + {m_o}{U_{oj}}{\bf{b}}}\\
{ + \sqrt {\frac{{2B({{\bf{X}}_{oj}}){\mu _{oj}}}}{{{m_o}}}} {{\widehat {\bf{v}}}_ {oj\bot} }}
\end{array}} \right)\cdot d{{\bf{X}}_{oj}}} \right)}\\
{ + \varepsilon_o{L_{{\bf{g}}_{oj}^{\bf{X}}}}\left( {\left( \begin{array}{l}
\frac{{{m_o}U_{oj}^2}}{2} + {\mu _{oj}}B({{\bf{X}}_{oj}})\\
 + \frac{{{q_o}\Phi \left( {{{\bf{X}}_{oj}},t} \right)}}{2\varepsilon_o}
\end{array} \right)dt} \right)}
\end{array}} \right]} } .
\end{eqnarray}

Eq.(\ref{ff1}) contains a summation like $\sum\limits_j {{\bf{g}}_{oj}^{\bf{X}}\cdot{\nabla _{oj}}\Phi \left( {{{\bf{X}}_{oj}},t} \right)}$, which can further be divided as the summation of different category ensemble, which is $\sum\limits_l {{\bf{g}}_{ol}^{\bf{X}}\cdot\nabla \Phi \left( {{\bf{X}},t} \right)} $. Here, subscript $l$ denotes the ensemble of charged particle all located at the guiding-center point $(\mathbf{X},\mu,U)$. Due to the homogeneous property, for any generator vector ${{\bf{g}}_{oh}^{\bf{X}}}$ for the particle $j$ in this ensemble, there always exists a particle denoted by subscript $k$ in this ensemble, the generator ${{\bf{g}}_{ok}^{\bf{X}}}$ of which equals $-{{\bf{g}}_{oh}^{\bf{X}}}$. Therefore, the following identity can be derived
\begin{equation}\label{fff8}
\sum\limits_j {{\bf{g}}_{oj}^{\bf{X}}\cdot\nabla \Phi \left( {{{\bf{X}}_{oj}},t} \right) = 0}.
\end{equation}
To calculate Eq.(\ref{ff1}), we also need another identity
\begin{equation}\label{fff9}
{\bf{g}}_{oj}^{\bf{X}}\cdot{\partial _\mu }\sqrt {\frac{{2B({{\bf{X}}_{oj}}){\mu _{oj}}}}{{{m_o}}}} {\widehat {\bf{v}}_ {oj\bot} }d\mu  = 0,
\end{equation}
which results from ${\bf{g}}_{oj}^{\bf{X}} \bot {\widehat {\bf{v}}_{oj \bot }}$.
Eventually, it's derived out that only terms like ${L_{{\bf{g}}_{oj}^{\bf{X}}}}\left( {\sqrt {\frac{{2B({{\bf{X}}_{oj}}){\mu _{oj}}}}{{{m_o}}}} {{\widehat {\bf{v}}}_{oj \bot }}\cdot d{{\bf{X}}_{oj}}} \right)$ in Eq.(\ref{ff1}) can generate non-zero results. The summation of these terms leads to
\begin{eqnarray}\label{ff4}
{\Gamma _1} && = \sum\limits_{o \in \{ i,e\} } {\sum\limits_j {{\varepsilon_o\bf{g}}_{oj}^{\bf{X}}\cdot{\partial _{\theta_{oj}} }\sqrt {\frac{{2B({{\bf{X}}_{oj}}){\mu _{oj}}}}{{{m_o}}}} {{\widehat {\bf{v}}}_ {oj\bot} }d\theta_{oj} } } \nonumber \\
&& = \sum\limits_{o \in \{ i,e\} } {\sum\limits_j {\varepsilon_o\frac{2 m_o\mu_{oj}}{q_o} d\theta_{oj} } }.
\end{eqnarray}

The next one is
\begin{equation}\label{fff5}
{\Gamma _2} = \frac{1}{2}\sum\limits_{o,n \in \{ i,e\} } {\sum\limits_{j,h} {{\varepsilon _o}{\varepsilon _n}{L_{{\bf{g}}_{oj}^{\bf{X}}}}{L_{{\bf{g}}_{nh}^{\bf{X}}}}\gamma \left( {\bf{Z}} \right)} }.
\end{equation}
We only keep the lower order part of $\Gamma_2$. The lower order  part of $\gamma$ is written as
\begin{equation}\label{ff6}
\Upsilon \left( {\bf{Z}} \right) = \sum\limits_{o \in \{ i,e\} } {\sum\limits_j {{\Upsilon _{oj}}\left( {\bf{Z}} \right)} },
\end{equation}
\begin{equation}\label{fff6}
{\Upsilon _{oj}}\left( {\bf{Z}} \right) = \frac{{{q_o}}}{{{\varepsilon _o}}}{\bf{A}}\left( {{{\bf{X}}_{oj}}} \right)\cdot d{{\bf{X}}_{oj}} - \frac{{{q_o}\phi \left( {{{\bf{X}}_{oj}},t} \right)}}{{2{\varepsilon _o}}}dt.
\end{equation}
To calculate Eq.(\ref{fff5}), it needs to be noted that only the terms with $n=o$ and $j=h$ in $\Gamma_2$ can produce nonzero terms based on the homogeneous assumption in the $\theta$ direction. The following formula is needed as well
\begin{equation}\label{fff7}
\begin{array}{l}
{\varepsilon _o}{q_o}L_{{\bf{g}}_{oj}^{\bf{X}}}^2\left( {{\bf{A}}\left( {{{\bf{X}}_{oj}}} \right)\cdot d{{\bf{X}}_{oj}}} \right)\\
 \approx  - {\varepsilon _o}{q_o}{\bf{g}}_{oj}^{\bf{X}}\cdot{\partial _{{\theta _{oj}}}}\left( {{\bf{g}}_{oj}^{\bf{X}} \times {\bf{B}}\left( {{{\bf{X}}_{oj}}} \right)} \right)d{\theta _{oj}}\\
 =  - {\varepsilon _o}\rho _0^2B\left( {{{\bf{X}}_{oj}}} \right)d\theta_{oj} =  - {\varepsilon _o}\frac{{2{m_o \mu_{oj}}}}{{{q_0}}}d{\theta _{oj}}.
\end{array}
\end{equation}
To derive the approximate equality in Eq.(\ref{fff7}), the formula $\varepsilon_o {\bf{g}}_{oj}^{\bf{X}} \times \nabla  \times \left( {{\bf{g}}_{oj}^{\bf{X}} \times {\bf{B}}\left( {{{\bf{X}}_{oj}}} \right)} \right)\cdot d{{\bf{X}}_{oj}}$ is neglected in the $\mathbf{X}_{oj}$ component in the Lagrangian 1-form, since its order is $O(\varepsilon_o)$, while the terms in the $\mathbf{X}_{oj}$ component in $\Gamma_0$ are of order $O(1/\varepsilon_o)$. In fact, this term can also be cancelled by introducing a generator $\mathbf{g}_{2oj}^{\mathbf{X}}$ of order $\varepsilon_o^2$ for each $j$. To derive the second equality in Eq.(\ref{fff7}), the equation ${\widehat {\bm{\rho }}_0} \times {\widehat {\bf{v}}_ \bot } =  - \widehat {\bf{b}}$ is adopted.

Eventually, the rest of $\Gamma_2$ is
\begin{eqnarray}\label{ff7}
{\Gamma _2} &&= \frac{1}{2}\sum\limits_{o \in \{ i,e\} } {\sum\limits_j {\varepsilon _o^2L_{{\bf{g}}_{oj}^{\bf{X}}}^2\Upsilon \left( {\bf{Z}} \right)} } \nonumber  \\
&& = \frac{1}{2}\sum\limits_{o \in \{ i,e\} } {\sum\limits_j {\left( { - {\varepsilon _o}\frac{{2{m_o}{\mu _{oj}}}}{{{q_o}}}d{\theta _{oj}} - {\varepsilon _o}{q_o}{{\left( {{\bf{g}}_{oj}^{\bf{X}}\cdot{\nabla _{oj}}} \right)}^2}\Phi \left( {{{\bf{X}}_{oj}},t} \right)dt} \right)} } \nonumber \\
&& = \frac{1}{2}\sum\limits_{o \in \{ i,e\} } {\sum\limits_j {\left( { - {\varepsilon _o}\frac{{2{m_o}{\mu _{oj}}}}{{{q_o}}}d{\theta _{oj}} - {\varepsilon _o}{q_o}\rho _{oj}^2\nabla _{oj}^2\Phi \left( {{{\bf{X}}_{oj}},t} \right)dt} \right)} }
\end{eqnarray}
In Eq.(\ref{ff7}), the factor $\frac{1}{2}$ before $\Phi(\mathbf{X}_{oj},t)$ is removed by combining all potential depending on $\mathbf{X}_{oj}$ together.
The third identity in Eq.(\ref{ff7}) also comes from the homogeneous assumption. At last, combining Eqs.(\ref{ff5},\ref{ff4},\ref{ff7}) together, we could derive the following fundamental Lagrangian 1-form defined on the new coordinates up to the second order approximation
\begin{equation}\label{ff8}
\Gamma  = \sum\limits_{o \in \{ i,e\} } {\sum\limits_j {\left[ {\begin{array}{*{20}{l}}
\begin{array}{l}
\left( {{q_o}{\bf{A}}\left( {{{\bf{X}}_{oj}}} \right) + {m_o}{U_{oj}}{\bf{b}}} \right)\cdot d{{\bf{X}}_{oj}}\\
 + \frac{{{m_o}{\mu _{oj}}}}{{{q_o}}}d{\theta _{oj}}
\end{array}\\
{ - \left( \begin{array}{l}
\frac{{{m_o}U_{oj}^2}}{2} + {\mu _{oj}}B({{\bf{X}}_{oj}})\\
 + {q_o}\frac{\Psi_o}{2} \left( {{{\bf{X}}_{oj}},{\mu _{oj}},t} \right)
\end{array} \right)dt}
\end{array}} \right]} }
\end{equation}
with
\begin{equation}\label{ff9}
\Psi_o \left( {{\bf{X}}_{oj},\mu_{oj},t} \right) =\Phi \left( {{{\bf{X}}_{oj}},t} \right) +\Pi_o \left( {{\bf{X}}_{oj},\mu_{oj},t} \right),
\end{equation}
\begin{equation}\label{ff10}
\Pi_o \left( {{\bf{X}}_{oj},\mu_{oj},t} \right)={{\rho _{oj}^2\nabla _{oj}^2\Phi \left( {{{\bf{X}}_{oj}},t} \right)}}.
\end{equation}
All the symbols $\varepsilon_o$ are removed from Eq.(\ref{ff8},\ref{ff9}). $\bar{\Gamma}$ is the new fundamental Lagrangian 1-form determining  the dynamics of all ions and electrons on the new coordinate system up to the second order approximation. $\Pi_o \left( {{\bf{X}}_{oj},\mu_{oj},t} \right)$ is a FLR term and could introduce the difference to the trajectory equations compared with those equations derived without $\Pi_o \left( {{\bf{X}}_{oj},\mu_{oj},t} \right)$.

\section{Modelling the fundamental Lagrangian 1-form on guiding-center coordinates by GCKM}\label{sec5}

\begin{figure}[htbp]
\centering
\includegraphics[height=4cm,width=5cm]{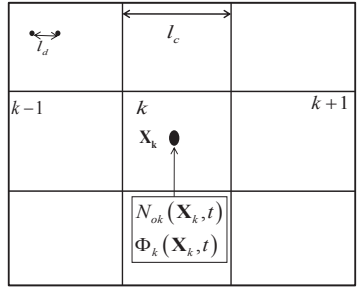}\centering
\caption{\label{gyrocell}  The schematic plot of cells and the location of $N_{ok}(\mathbf{X}_k,t)$ and $\Phi_{k}(\mathbf{X}_k,t)$ with $l_d\ll l_c \ll l_p$.}
\end{figure}

Eq.(\ref{ff8}) is a model based on first-principle force. It becomes untractable by increasing the particle number $N$. A way to simplify Eq.(\ref{ff8}) is to adopt the same modeling method used in Sec.(\ref{sec3}) by degenerating the pair-wise Coulomb force to the Poisson equation plus a boundary condition. The potential $\Phi \left( {{{\bf{X}}_{oj}}},t \right)$ experienced by the particle located at $\mathbf{X}_{oj}$ is given by Eq.(\ref{ef24}). The knowledge of $\Pi_o \left( {{\bf{X}}_{oj},\mu_{oj},t} \right)$  depends on $\Phi \left( {{{\bf{X}}_{oj}}},t \right)$. The formalism of $\Phi \left( {{{\bf{X}}_{oj}}},t \right)$ makes sure that we can use Poisson's equation plus a boundary condition to model it. Similar to Eq.(\ref{vp58}), the relation between $\Phi \left( {{{\bf{X}}_{oj}}},t \right)$ and the local charge density can be written as
\begin{equation}\label{ff11}
{\nabla ^2}{\Phi _k}({{\bf{X}}_k},t) = \frac{1}{{{\epsilon_0}}}\left( {e{N_{ek}}({{\bf{X}}_k},t) - q{N_{ik}}({{\bf{X}}_k},t)} \right).
\end{equation}
${{N_{ek}}({{\bf{X}}_k},t)}$ and ${{N_{ik}}({{\bf{X}}_k},t)}$ are the average density of electrons and ions of the $k$th cell on the new coordinates and located at the spatial center of the $k$th cell. $\Phi _k({{\bf{X}}_k},t)$ is the average electrostatic potential at the center of the $k$th cell. The sketch map is given in Fig.(\ref{gyrocell})

Just as Eq.(\ref{ef5}), with the modeled electrostatic potential $\Phi_k(\mathbf{X}_k,t)$, the dynamics of the $j$th particle in the $k$th cell on the new coordinate can be described by the following test Lagrangian 1-form
\begin{equation}\label{ff12}
\begin{array}{*{20}{l}}
{{\Gamma _{okj}} = \left( {{q_o}{\bf{A}}\left( {{{\bf{X}}_{oj}}} \right) + {m_o}{U_{oj}}{\bf{b}}} \right)\cdot d{{\bf{X}}_{oj}}}+\frac{m_o \mu_{oj}}{q_o}d\theta_{oj}\\
{ - \left( {{\mu _o}B\left( {{{\bf{X}}_{oj}}} \right) + \frac{{{m_o}U_{oj}^2}}{2} + {q_o}{\bar{\Psi} _{ok}}({{\bf{X}}_k},{\mu _{oj}},t)} \right)dt,}
\end{array}
\end{equation}
with
\begin{equation}\label{ff14}
\bar{\Psi}_{ok} \left( {{{\bf{X}}_{k}},{\mu _{oj}},t} \right) ={\Phi _k}\left( {{{\bf{X}}_k},t} \right) +\frac{{\Pi_{ok} \left( {{{\bf{X}}_{k}},{\mu _{oj}},t} \right)}}{2},
\end{equation}
\begin{equation}\label{ff13}
{\Pi _{ok}}\left( {{{\bf{X}}_k},{\mu _{oj}},t} \right) = \rho _{oj}^2{\nabla ^2}\Phi_k \left( {{{\bf{X}}_k},t} \right),
\end{equation}
where the second order derivative ${\nabla ^2}\Phi_k \left( {{{\bf{X}}_k},t} \right)$  is given by a middle-point discrete derivative with $\mathbf{X}_k$ as the center.
Here, the factor $\frac{1}{2}$ before the electrostatic potential is removed, since all the mutual interactions depending on $\mathbf{X}_{oj}$ are combined together and is approximated by ${\Phi _k}\left( {{{\bf{X}}_k},t} \right)$.
Eq.(\ref{ff12}) determines the trajectory equations on the new coordinates. ${\Pi _{ok}}\left( {{{\bf{X}}_k},{\mu _{oj}},t} \right)$ denotes the FLR term introduced to the trajectory equations.

Now, we need to calculate the evolution of the density of charged particles in the $k$th cell. To do this, the Klimontovich distribution on the  coordinate frame $\mathbf{Z}'=\{\mathbf{X},\mu, U\}$ is needed
\begin{eqnarray}\label{ef28}
&& \mathbb{M}{_{ok}}\left( {\mathbf{Z}'} \right) \nonumber \\
&& = \sum\limits_j {\frac{{\delta \left( {{\bf{X}} - {{\bf{X}}_{oj}(t)}} \right)\delta \left( {\mu  - {\mu _{oj}}(t)} \right)\delta \left( {U - {U_{oj}}(t)} \right)}}{{B\left( {{{\bf{X}}_{oj}}}(t) \right)}}}.
\end{eqnarray}
The independence of $\theta$ is naturally inherited by $\mathbb{M}{_{ok}}\left( {\mathbf{Z}'} \right)$ from the fundamental Lagrangian 1-form Eq.(\ref{ff8}) which is  independent of $\theta$.

The evolution of $\mathbb{M}{_{ok}}\left( \mathbf{Z}' \right)$ is given by the Liouville's equation
\begin{equation}\label{ef29}
\left( {\frac{\partial }{{\partial t}} + \frac{{d{{\bf{X}}}}}{{dt}}\cdot\nabla  + \frac{{d{U}}}{{dt}}\frac{\partial }{{\partial U}}} \right)\mathbb{M}{_{ok}}(\mathbf{Z}') = 0
\end{equation}
By ensemble summation of $\mathbb{M}{_{ok}}$, the Vlasov distribution ${F}_{o}(\mathbf{Z}')$ can be derived and Eq.(\ref{ef29}) becomes the Vlasov equation
\begin{equation}\label{ff15}
\left( {\frac{\partial }{{\partial t}} + \frac{{d{\bf{X}}}}{{dt}}\cdot\nabla  + \frac{{dU}}{{dt}}\frac{\partial }{{\partial U}}} \right){F_{ok}}({\mathbf{Z}'}) = 0.
\end{equation}

By shrinking $l_c$ to be small enough (much smaller than the Larmor radius of ions), the subscript $k$ can be removed from Poisson's equation, the Lagrangian 1-form of a test particle, and the Vlasov equation, all of which are rewritten as
\begin{equation}\label{ff16}
{\nabla ^2}\Phi ({\bf{X}},t) = \frac{1}{{{\epsilon_0}}}\left( {e{N_e}({\bf{X}},t) - q{N_i}({\bf{X}},t)} \right),
\end{equation}
\begin{equation}\label{ff17}
\begin{array}{*{20}{l}}
{{\Gamma _{o}} = \left( {{q_o}{\bf{A}}\left( {{{\bf{X}}}} \right) + {m_o}{U}{\bf{b}}} \right)\cdot d{{\bf{X}}}}+\frac{m_o\mu}{q_o}d\theta\\
{ - \left( {{\mu}B\left( {{{\bf{X}}}} \right) + \frac{{{m}U^2}}{2} + {q_o}\bar{\Psi}_o ({\bf{X}},{\mu},t)} \right)dt,}
\end{array}
\end{equation}
\begin{equation}\label{ff18}
\left( {\frac{\partial }{{\partial t}} + \frac{{d{\bf{X}}}}{{dt}}\cdot\nabla  + \frac{{dU}}{{dt}}\frac{\partial }{{\partial U}}} \right){F_o}({\mathbf{Z}'}) = 0,
\end{equation}
with
\begin{equation}\label{ff19}
\bar{\Psi}_o \left( {{\bf{X}},{\mu},t} \right) =\Phi \left( {{\bf{X}},t} \right) +\frac{\Pi_o \left( {{\bf{X}},{\mu},t} \right)}{2},
\end{equation}
\begin{equation}\label{ff20}
\Pi_o \left( {{\bf{X}},{\mu},t} \right) ={\rho_o(\mathbf{X,\mu})^2{\nabla ^2}\Phi \left( {{\bf{X}},t} \right)}.
\end{equation}

So far, we derived a guiding-center Klimontovich-Poisson model comprising Eqs.(\ref{ff16},\ref{ff17},\ref{ef28})
and a guiding-center Vlasov-Poisson model comprising Eqs.(\ref{ff16}-\ref{ff18}) as the modeling of the fundamental Lagrangian 1-form in Eq.(\ref{ff8}). These two models constitute a close system, respectively, based on which all theoretical analysis and simulations can be done.

%


\section{The equation of quasi-neutral condition in GCKM}\label{sec6}

The new guiding-center Vlasov-Poisson model is taken as the example to show the equation for the quasi-neutral condition on guiding-center coordiantes. In this paper, the electrostatic potential with $l_p\gg \rho_e$ while $l_p \approx \rho_i$ is taken into account. The FLR term for electrons in Eq.(\ref{ff20}) denoted  by $\Pi \left( {{\bf{X}},{\mu _{ej}},t} \right) $ can be ignored, due to the much smaller Larmor radius of electrons. The adiabatic approximation of the fluctuation density of electrons can be written as
\begin{equation}\label{vp47}
{{N}_e}({\bf{X}},t) = {{N}_{e0}}({\bf{X}}) + \frac{{e \Phi \left( {{\bf{X}}},t \right)}}{{{T_e}}}{{N}_{e0}}({\bf{X}}),
\end{equation}
 with ${N}_{e0}(\bf{X})$ being the equilibrium density for electrons on the new coordinates. The FLR term denoted by $\Pi \left( {{\bf{X}},{\mu _{oj}},t} \right)$ for ions is kept. The distribution function of ions on new coordinates can be decomposed as \cite{1980antonsen,2014shuangxizhang}
\begin{equation}\label{ff21}
{F_i}({\mathbf{Z}'}) = {F_{i0}}({\bf{Z}}') + {F_{i1}}\left( {\mathbf{Z}'} \right).
\end{equation}
with the perturbed part being
\begin{equation}\label{ff22}
{F_{i1}}\left(\mathbf{Z}' \right) =  - \frac{{q\Psi \left( {{\bf{X}},{\mu,t}} \right)}}{{{T_i}}}{F_{i0}}({\mathbf{Z}'}) + H_i\left( \mathbf{Z}' \right).
\end{equation}
Here, $H\left( \mathbf{Z} \right)$ is the non-adiabatic distribution. The perturbed density is derived by integrating over the distribution function
\begin{equation}\label{ff23}
\begin{array}{l}
{N_{i1}}\left( {{\bf{X}},t} \right) = \int {{F_{i1}}\left( \mathbf{Z}' \right)B\left( {\bf{X}} \right){d^3}{\bf{X}}d\mu dU} \\
 =  - \frac{{q{N_{i0}}\left( {{\bf{X}},t} \right)}}{{{T_i}}}\left( {1 + \rho _t^2{\nabla ^2}} \right)\Phi \left( {{\bf{X}},t} \right) \\
 + q\int {H\left( \mathbf{Z}'\right)B\left( {\bf{X}} \right){d^3}{\bf{X}}d\mu dU},
\end{array}
\end{equation}
where $\rho_t$ is the Larmor radius with the thermal velocity.
The equation for the quasi-neutral condition becomes
\begin{equation}\label{ff24}
\begin{array}{l}
\frac{{{e^2}\Phi \left( {{\bf{X}},t} \right)N_{e0}(\mathbf{X})}}{{{T_e}}} =  - \frac{{{q^2}{N_{i0}}\left( {{\bf{X}}} \right)}}{{{T_i}}}\left( {1 + \rho _t^2{\nabla ^2}} \right)\Phi \left( {{\bf{X}},t} \right)\\
 + q\int {H\left( \mathbf{Z}' \right)B\left( {\bf{X}} \right){d^3}{\bf{X}}d\mu dU}.
\end{array}
\end{equation}
The term proportional to $\rho _t^2{\nabla ^2}$ is a kind of the polarization density as explained by Eq.(38) of Ref.(\cite{2007brizard1}). But for the practical application, the edition of the equation for the quasi-neutral condition is
\begin{equation}\label{ff28}
\begin{array}{l}
\frac{{{e^2}{N_{e0}}\left( {{\bf{X}},t} \right)\Phi \left( {{\bf{X}},t} \right)}}{{{T_e}}}  \\
 = q\int {{F_{i1}}\left( \mathbf{Z}' \right)B\left( {\bf{X}} \right){d^3}{\bf{X}}d\mu dU}.
\end{array}
\end{equation}
For the numerical application of Eq.(\ref{ff28}), $F_{i1}$ can be straightforwardly calculated from the Vlasov equation based on the potential given by the last time step. To get the solution of the new potential on the current step, it's not needed to solve operator ``$\nabla^2$" which appears in the equation for the quasi-neutral equation in CGM.


\section{Applications of GCKM}\label{sec7}

For numerical applications of GCKM, the simulations are carried out totally on the new coordinate. Except at the beginning and at the end of the simulations, the calculations of Fourier spectrum of the perturbations and the average of the gyroangle are exempted at each time step. So the numerical time  and numerical instabilities can be significantly reduced.


For the simulation based the Vlasov distribution, the initial distribution of ${{F}_o}\left( {{\bf{X}},\mu ,U,t} \right)$ needs to be transformed from the one on particle's coordinates. If we know an initial distribution function $f_{oin}(\mathbf{x},u_1,\mu_1,t)$ for ions on particle's coordinates, the initial distribution on $\mathbf{Z}'$ is given by
\begin{eqnarray}\label{vp38}
&& {F_{iin}}\left( {\bf{Z}}' \right)  \nonumber \\
&& = \int{ \begin{array}{*{20}{l}}
{{f_{iin}}\left( {\bf{z}} \right)\delta \left( {{\bf{x}} - {\bf{X}} - \bm{\rho}_0 \left( {{\bf{X}},\mu ,u,\theta } \right)} \right)}\\
{ \times \delta \left( {U - {\bf{v}}\cdot{\bf{b}}} \right)\delta \left( {\mu  - \frac{{{m_i}v_ \bot ^2}}{{2B\left( {\bf{X}} \right)}}} \right)\frac{{{d^3}{\bf{x}}d{\bf{v}}d\theta }}{{2\pi B\left( {\bf{X}} \right)}}}
\end{array}}
\end{eqnarray}
In fact, Eq.(\ref{vp38}) is the first order approximation of Eq.(\ref{ff26}).
The transform given by Eq.(\ref{vp38}) naturally makes ${F}_{iin}$ inherit the perturbation wave  from $f_{iin}$.
Using the initial distribution $F_{iin}\left( \mathbf{Z} \right)$, Poisson's equation Eq.(\ref{ff16}), and trajectory equations derived from Eq.(\ref{ff17}), the time evolution of the distribution in Eq.(\ref{ff18}) can be calculated to get $F_{iend}\left(\mathbf{Z} \right)$, based on which various quantities, e.g, transport of number density, momentum, energy, and amplitude of potential fluctuation,  can be derived on new coordinate $\bf{Z}$. ${F}_{iend}\left( \mathbf{Z}  \right)$ can also be transformed back to the one on particle's coordinates by the following formula
\begin{eqnarray}\label{vp41}
{f_{iend}}\left( {\bf{z}} \right) = \int{ \begin{array}{l}
{F_{iend}}\left( {\bf{Z}}' \right)\delta \left( {{\bf{x}} - {\bf{X}} - \bm{\rho}_0 \left( {{\bf{X}},\mu ,U,\theta } \right)} \right)\\
 \times \delta \left( {U - {\bf{v}}\cdot{\bf{b}}} \right)\delta \left( {\mu  - \frac{{{m_i}v_ \bot ^2}}{{2B\left( {\bf{X}} \right)}}} \right)\\
\frac{{{d^3}{\bf{X}}d\mu dUd\theta }}{{2\pi}}
\end{array} }
\end{eqnarray}
The perturbation wave included by $\mathbb{F}_{iend}(\mathbf{Z})$ is naturally transmitted  back to $f_{iend}(\mathbf{z})$ by Eq.(\ref{vp41}).

For particle-in-cell simulations, all equations needed are the trajectory equations derived from Eq.(\ref{ff17}), the Poisson's equation given by Eq.(\ref{ff16}) and the evolution equation Eq.(\ref{ef29}) for the Klimontovich distribution. The four-point scheme to calculate the density on particle coordinate from the one on the gyrocenter coordinate is exempted at each time step. Eqs.(\ref{vp38}) and (\ref{vp41}) may be applied to transform the distribution between $\bf{z}$ and $\bf{Z}$ at the beginning and at the end of the simulation.

\section{Summary and discussion}\label{sec8}

This paper presented an assumption of a homogeneous distribution of the ensemble of charged particles over the gyrophase. As a company to this assumption, this paper developed a fundamental Lagrangian 1-form, which contains all particles in the magnetized plasma system as well as the Coulomb pair force between particles instead of field-particle interaction used in conventional gyrokinetic models. This fundamental Lagrangian 1-form perfectly makes use of the property of the homogeneous assumption to reduce the gyrophase of each particle. Therefore, it doesn't need an additional gyrocenter transform as used in CGM, which induces a polarization density in the quasi-neutral equation in CGM. However, such a polarization density doesn't appear in GCKM. It's expected that the numerical application of GCKM could reduce the numerical noise level and guarantee long-term simulation.

In this paper, the fundamental Lagrangian 1-form is only approximated up to the second order $O(\varepsilon_o^2)$. High order approximation can be carried out for specific problems. This paper only considers electrostatic perturbations, whilst the magnetic vector potential is treated as the background field. For  magnetic perturbations, such an operation is improper. The non-equilibrium part of the magnetic vector potential needs special treatment similar to the treatment of the electrostatic potential in this paper. This is left for the future work.

\section{Acknowledgments}\label{sec9}

This work was completed at Uji campus of Kyoto University, Japan. The author is indebted to the discussion with
Prof. Weixing Wang, Prof. Yasuaki Kishimoto,Prof. Kenji Imadera, Prof. Alain Brizard, Prof. T.S.Hahm, Prof. Guoyong Fu,
in particular, is grateful to Prof. Johan Anderson for reading the manuscript.

\appendix

\section{Background field, non-background field and scalar under the coordinate transform}\label{sec10}

The background field is given by the magnetic vector potential $\mathbf{A}(\mathbf{x})$, which doesn't explicitly dependent on time.  The other field is the non-background field which is the electrostatic potential generated by the separation of ions and electrons.
It should be noted that $\Phi(\mathbf{X},t)$ given by Eq.(\ref{ef24}) is different from $\phi(\mathbf{x},t)$ given by Eq.(\ref{ef131}) in the following way. $\phi(\mathbf{x},t)$ and $\Phi(\mathbf{X},t)$ both depend on the spatial coordinate of all particles in the respective spatial space. The coordinate transform which induces the pullback transform Eq.(\ref{ef21}) is approximated as
\begin{equation}\label{ef26}
{{\bf{x}}_{oj}} \approx {{\bf{X}}_{oj}} - {\bf{g}}_{oj}^{{{\bf{X}}}}.
\end{equation}
${\bf{g}}_{oj}^{{{\bf{X}}}}$ depends on $\mu_{oj}$. If  the RHS of Eq.(\ref{ef26}) is substituted back to $\phi(\mathbf{x},t)$, it's found that $\phi(\mathbf{x},t)$ depends on $(\mathbf{X}_{oj},\mu_{oj})$ for all $j$ and $o\in\{i,e\}$ except the new coordinate of this particle. However, $\Phi(\mathbf{X},t)$ only depends on $\mathbf{X}_{oj}$ for all $o$ and $j$,  not on any $\mu_{oj}$.
This also explains why $\phi(\mathbf{x},t)$ is not a scalar under coordinate transform $\psi$.


The fact that $\phi(\mathbf{x},t)$ is not a scalar can also be explained as follows.
As introduced in Sec.(\ref{sec1}), the coordinate transform $\psi$ is defined on the phase space as $\psi: \mathbf{z}\equiv(\mathbf{x},\mathbf{v})\to\mathbf{Z}\equiv (\mathbf{X},\mu, U,\theta)$. The Cary-Littlejohn single-parameter Lie transform theory\cite{1983cary} shows that for a coordinate transform described by a group of autonomous equations like
\begin{equation}\label{ef8}
\frac{{d Z_f^i}}{{d \varepsilon }}\left( {{\bf{z}},\varepsilon } \right) = {g^i}\left( {\bf{Z}} \right),
\end{equation}
and
\begin{equation}\label{ef9}
\frac{{d{\bf{z}}}}{{d\varepsilon }} = 0,
\end{equation}
if there exists a scalar $s(\bf{z})$ with its counterpart in the new coordinate transform being $S(\mathbf{Z})$, then, from the identity $s\left( {\bf{z}} \right) = S\left( {{\bf{Z}},\varepsilon } \right)$, the following formula can be derived
\begin{equation}\label{ef10}
\frac{{\partial S}}{{\partial \varepsilon }}\left( {{\bf{Z}},\varepsilon } \right) =  - {g^\mu }\left( {\bf{Z}} \right)\frac{{\partial S}}{{\partial {Z^\mu }}}\left( {{\bf{Z}},\varepsilon } \right).
\end{equation}

Now, if we assume that $\phi(\mathbf{x},t)$ is a scalar under the coordinate transform, and its new form is denoted as $\Phi(\mathbf{X},t,\varepsilon)$ in the new coordinate, similar to Eq.(\ref{ef10}), we could derive an equation
\begin{equation}\label{ef11}
\frac{{\partial \Phi \left( {{\bf{X}},t,\varepsilon } \right)}}{{\partial \varepsilon }} =  - {g^\mu }\left( {\bf{Z}} \right)\frac{{\partial \Phi \left( {{\bf{X}},t,\varepsilon } \right)}}{{\partial {Z^\mu }}}.
\end{equation}
In Eq.(\ref{ef11}), ${g^\mu }\left( {\bf{Z}} \right)$ is a function of variables $(\mathbf{X},\mu,\theta)$. But $\Phi(\bf{X},t)$ is only a function over  $\mathbf{X},t$. Therefore, Eq.(\ref{ef11}) is not valid. The essential reason is that $\phi(\bf{x},t)$ is not a scalar under coordinate transform.

However, the distribution is a scalar, the transform of which is
\begin{equation}\label{ff25}
\frac{{\partial {F_o}\left( {\bf{Z}} \right)}}{{\partial \varepsilon }} =  - {g^\mu }\left( {\bf{Z}} \right)\frac{{\partial {F_o}\left( {\bf{Z}} \right)}}{{\partial {Z^\mu }}}.
\end{equation}
The solution of Eq.(\ref{ff25}) is
\begin{equation}\label{ff26}
{F_o}\left( {\bf{Z}} \right){\rm{ = exp}}\left( { - {\bf{g}}\left( {\bf{Z}} \right)\cdot\nabla } \right){f_o}\left( {\bf{Z}} \right).
\end{equation}

\section{Energy conservation of the system given by Eq.(\ref{ff8})}\label{ap3}  

The trajectory equations derived from Eq.(\ref{ff8}) are
\begin{equation}\label{gf1}
\dot{\mathbf{X}}_{oj}{\rm{ = }}\frac{{{U_{oj}}{\bf{B}}_{oj}^* + {\bf{b}} \times \nabla {H_{oj}}}}{{{\bf{b}}\cdot{\bf{B}}_{oj}^*}}
\end{equation}
\begin{equation}\label{gf2}
{{\dot U}_{oj}} = \frac{{ - {\bf{B}}_{oj}^*\cdot\nabla {H_{oj}}}}{{{\bf{b}}\cdot{\bf{B}}_{oj}^*}},
\end{equation}
\begin{equation}\label{gf3}
{{\dot \mu }_{oj}} = 0,
\end{equation}
with ${\bf{B}}_{oj}^* = {\nabla _{oj}} \times \left( {{\bf{A}}\left( {{{\bf{X}}_{oj}}} \right) + {U_{oj}}{\bf{b}}} \right)$.
The energy per particle is formally written as
\begin{equation}\label{gf4}
{H_{oj}} = \frac{{{m_o}U_{oj}^2}}{2} + {\mu _{oj}}B({{\bf{X}}_{oj}}) + {q_o}\bar{\Psi} \left( {{{\bf{X}}_{oj}},{\mu _{oj}}} \right).
\end{equation}
Now, we try to prove $\frac{d}{{dt}}\sum\limits_{o,j} {{H_{oj}}}  = 0$. It's first to derive the following three identities
\begin{equation}\label{gf5}
\begin{array}{*{20}{l}}
{\sum\limits_{o,j} {{m_o}\frac{d}{{dt}}U_{oj}^2}  = \sum\limits_{o,j} {\frac{{ - {U_{oj}}{\bf{B}}_{oj}^*\cdot{\nabla _{oj}}\sum\limits_{n \in \{ i,e\} ,h} {{H_{nh}}} }}{{{\bf{b}}\cdot{\bf{B}}_{oj}^*}}} }\\
{ = \sum\limits_{o,j} {\frac{{ - {U_{oj}}{\bf{B}}_{oj}^*\cdot{\nabla _{oj}}\sum\limits_{n \in \{ i,e\} ,h} {\left( \begin{array}{l}
{m_n}U_{nh}^2 + \overbrace {{\mu _{nh}}B\left( {{{\bf{X}}_{nh}}} \right)}^{(2.1)}\\
 + \overbrace {{q_n}{\Psi _n}\left( {{{\bf{X}}_{nh}},{\mu _{nh}}} \right)}^{(1.1)}
\end{array} \right)} }}{{{\bf{b}}\cdot{\bf{B}}_{oj}^*}}} ,}
\end{array}
\end{equation}
\begin{equation}\label{gf6}
\begin{array}{*{20}{l}}
{\sum\limits_{o,j} {{\mu _o}\frac{{dB\left( {{{\bf{X}}_{oj}}} \right)}}{{dt}} = } \sum\limits_{o,j} {{\mu _o}\frac{{d{{\bf{X}}_{oj}}}}{{dt}}\cdot{\nabla _{oj}}B\left( {{{\bf{X}}_{oj}}} \right)} }\\
{ = \sum\limits_{o,j} {\frac{{\left( \begin{array}{l}
\overbrace {{\mu _{oj}}{U_{oj}}{\bf{B}}_{oj}^*}^{(2.2)}\\
 + \overbrace {{\mu _{oj}}{\bf{b}} \times {\nabla _{oj}}\sum\limits_{n \in \{ i,e\} ,h} {{H_{nh}}} }^{(3.1)}
\end{array} \right)}}{{{\bf{b}}\cdot{\bf{B}}_{oj}^*}}\cdot{\nabla _{oj}}B\left( {{{\bf{X}}_{oj}}} \right)} ,}
\end{array}
\end{equation}
\begin{equation}\label{gf7}
\begin{array}{*{20}{l}}
\begin{array}{l}
\frac{d}{{dt}}\sum\limits_{oj} {{q_o}\Psi \left( {{{\bf{X}}_{oj}},{\mu _{oj}}} \right)} \\
 = \sum\limits_{n \in \{ i,e\} ,h} {\frac{{d{{\bf{X}}_{nh}}}}{{dt}}\cdot{\nabla _{nh}}{q_o}\sum\limits_{o \in \{ i,e\} ,j} {\Psi \left( {{{\bf{X}}_{oj}},{\mu _{oj}}} \right)} }
\end{array}\\
{ = \sum\limits_{n,h} {\left( \begin{array}{l}
\frac{{\left( {\overbrace {{U_{nh}}{\bf{B}}_{nh}^*}^{(1.2)} + \overbrace {{\bf{b}} \times {\nabla _{nh}}\sum\limits_{m \in \{ i,e\} ,k} {{H_{mk}}} }^{(3.2)}} \right)}}{{{\bf{b}}\cdot{\bf{B}}_{oj}^*}}\\
\cdot{\nabla _{nh}}{q_o}\sum\limits_{o \in \{ i,e\} ,j} {\Psi \left( {{{\bf{X}}_{oj}},{\mu _{oj}}} \right)}
\end{array} \right)} .}
\end{array}
\end{equation}
In Eq.(\ref{gf7}), the property that ${\Psi \left( {{{\bf{X}}_{oj}},{\mu _{oj}}} \right)}$ doesn't explicitly depend on the time is applied. This is basic property of the electrostatic fluctuation. It's easy to observed that term (1.1) cancels (1.2), term (2.1) cancels (2.2). Term (3.1) cancels (3.2) based on the following two identities
\begin{equation}\label{gf8}
\begin{array}{l}
{\mu _{oj}}{\bf{b}} \times {\nabla _{oj}}{H_{oj}}\cdot{\nabla _{oj}}B\left( {{{\bf{X}}_{oj}}} \right)\\
 = {\mu _{oj}}{\bf{b}} \times {\nabla _{oj}}{q_o}\Psi \left( {{{\bf{X}}_{oj}},{\mu _{oj}}} \right)\cdot{\nabla _{oj}}B\left( {{{\bf{X}}_{oj}}} \right),
\end{array}
\end{equation}
\begin{equation}\label{gf9}
\begin{array}{l}
{\bf{b}} \times {\nabla _{oj}}{H_{oj}}\cdot{\nabla _{oj}}{q_o}\Psi \left( {{{\bf{X}}_{oj}},{\mu _{oj}}} \right)\\
 = {\mu _{oj}}{\bf{b}} \times {\nabla _{oj}}B\left( {{{\bf{X}}_{oj}}} \right)\cdot{\nabla _{oj}}{q_o}\Psi \left( {{{\bf{X}}_{oj}},{\mu _{oj}}} \right)
\end{array}.
\end{equation}
Then, the conservation of the total energy of the system is proved.

\newpage
\section*{References}

\bibliographystyle{pst}
\bibliography{NGVP}

\end{document}